\documentclass[epj]{svjour}

\usepackage{graphicx}
\usepackage{fancyhdr}

\def\mytitle{My title} 
\def\myauthors{My name}  
\def\mytype{My type of session}
\def\mysession{My session}

\def\mytitle{Double Parton Scattering in Associate Higgs Production} 
\def\myauthors{M. Y. Hussein}    
\def\mytype{Contributed Talk}
\def\mysession{Colliders - Higgs Phenomenology}

\begin{document}
\title{Double Parton Scattering in Associate Higgs Boson Production with Bottom Quarks at Hadron Colliders}

\author{M. Y. Hussein\inst{1}
\thanks{\emph{Email:} mhussein@sci.uob.bh}%
}                     

\institute{Department of Physics, College of Science,
  University of Bahrain, P.O. Box 32038, Kingdom of Bahrain}
%
\date{}
\abstract{
Higgs boson production in association with bottom quarks, is one of
the most important discovery channels for Higgs particles in the
Standard Model (SM) and its supersymmetric extension at the LHC pp
collider . The theoretical prediction of the corresponding cross
section has been improved by including the complete next-to-leading
order QCD corrections. We review the status results for the leading
order to the partonic process and present calculations for the
integrated cross-section in both single and double parton scattering
collisions.
} 
\maketitle
\section{Introduction}
\label{intro}

The discovery of the SM Higgs boson is one of the most important goals
and pressing issues of present and future colliders. An important
prerequisite for identifying the most convenient signatures for
detecting Higgs boson needs a precise knowledge of the various
production cross-sections, decay branching ratios, decay width, and
their couplings to other particles.

Recently, much progress has been made in the detection of a Higgs
boson, next-leading order corrections are now known for most
sub-process~\cite{Han91,Dit04,Dws91,Djo91,Grn93,Spr91}, knowledge of
parton distribution functions has improved as more deep inelastic data
become available and the range of possible input parameter values
decreased.

The dominant production of a SM Higgs boson in hadronic interactions
is through gluon-gluon fusion~\cite{Mch05,Han92,Cmp03}. Various
channels can be explored to search for Higgs boson at hadron
colliders. There are only a few Higgs production mechanism which lead
to detectable cross-section. Each of them use the preference of
coupling of the SM Higgs to heavy particles either massive vector
bosons or massive quarks.

The associated production of a Higgs boson with a pair of $b\bar{b}$
quarks has a small cross-section which is due to small size of Yukawa
coupling $g_{b\bar{b}h} = m_b/v \simeq 0.02$ in the
SM~\cite{Dws05}. In some extensions of the SM, such as the MSSM, the
Yukawa coupling of b-quarks can become strongly enhanced, the
associate production of a Higgs boson with a pair of $b\bar{b}$ quarks
can dominate over other production channels and this production
mechanism can be a significant source of Higgs bosons~\cite{Dws04}.

Detecting two bottom quarks in the final state identifies uniquely the
Higgs coupling responsible for the enhanced cross-section and
drastically reduces the background. This corresponds to an experiment
measuring the Higgs decay along two high pt bottom quark jets.

In a four-flavor-number scheme with no b-quarks in the initial
state, the lowest order processes are the tree level contributions
from \mbox{\(gg \rightarrow b\bar{b}h\)} and \mbox{\(q\bar{q} \rightarrow
b\bar{b}h\)}.

At high energies and due to large flux in particular at the LHC,
another type of scattering mechanism contributes to the cross section
besides to single scattering. Thus, for \(q\bar{q} \rightarrow
b\bar{b}h\) production there would be two computing mechanisms: single
parton scattering and double parton scattering featuring two Drell-Yan
processes happening simultaneously.

The purpose of the present work is to point out that the same
$b\bar{b}h$ final state can be produced also by double parton
scattering collision process.  The large rate of production of
$b\bar{b}$ pairs expected at the LHC gives rise to a relatively large
probability of production of a $b\bar{b}h$ in the process underling
the H production. In fact as a result of the present analysis it is
found that double parton scattering may represent a rather sizable
source of background.


\section{Double Parton Scattering Mechanism}

The multiple parton scattering occurs when two or more different pairs
of parton scatter independently in the same hadronic collision
\cite{Cmp04,Lnd87,Dlf00}.
            
With the only assumption of factorization of the two hard parton
processes A and B, the inclusive cross section of a double
parton-scattering in a hadronic collision is expressed by:

\begin{eqnarray}
  \lefteqn{\sigma_{(A,B)}^D = \frac{m}{2} \sum_{i,j,k,l}
    \Gamma_{ij}(x_1,x_2;b) \hat{\sigma}^A_{ik}(x_1,x'_1) } \nonumber
    \\ & & \hat{\sigma}^B_{jl}(x_2,x'_2) \Gamma_{kl}(x'_1,x'_2;b) dx_1
    dx'_1 dx_2 dx'_2 d^2b
\end{eqnarray}

Where $\Gamma_{ij}(x_1,x_2;b)$ are the double parton distribution
function, depending on the fractional momenta $x_1$, $x_2$ and the
relative transverse distance $b$ of the two parton undergoing the hard
processes A and B, the indices $i$ and $j$ refer to the different
parton species and $\hat{\sigma}^A_{ik}$ and $\hat{\sigma}^B_{jl}$ are
the partonic cross section.  The factor $m/2$ is for symmetry,
specifically $m=1$ for indistinguishable parton processes and $m=2$
for distinguishable processes.

The double distributions $\Gamma_{ij}(x_1,x_2;b)$ are the main reason
of interest in multiparton collisions. This distribution contains in
fact all the information of probing the hadron in two different points
contemporarily through the hard processes A and B.  The cross section
for multiparton process is sizable when the flux of partons is large,
namely at small $x$. Given the large flux one may hence expect that
correlations in momentum fraction will not be a major effect and
partons to be rather correlated in transverse space. Neglecting the
effect of parton correlations in $x$ one writes:

\begin{equation}
\Gamma_{ij}(x_1,x_2;b) = \Gamma_i(x_1) \Gamma_j(x_2) F^i_j(b)
\end{equation}

where $\Gamma_i(x) $the usual one body parton distribution $F^i_j(b)$
is function and is a function normalized to one and representing the
pair density in transverse space. The inclusive cross section hence
simplifies to:

\begin{equation}
  \sigma_{(A,B)}^D = \frac{m}{2} \sum_{ijkl} \Theta^{ij}_{kl}\hat{\sigma}_{jl}(A)\hat{\sigma}_{kl}(B)
\end{equation}
where $\hat{\sigma}_{jl}(A)$ and $\hat{\sigma}_{kl}(B)$ are the
hadronic inclusive cross section for the two partons labeled $i$ and
$j$ undergoes the hard interaction labeled A and for two partons $k$
and $l$ to undergo the hard interaction labeled B;
\begin{equation}
  \Theta^{ij}_{kl} = \int d^2 b F^i_k(b) F^j_l(b)
\end{equation}
are the geometrical coefficients with dimension an inverse cross
section and depending on various parton processes. These coefficients
are the experimentally accessible quantities carrying the information
of the parton correlation in transverse momentum. The cross section
for multiple parton collisions has been further simplified as:
\begin{equation}
  \sigma_{(A,B)}^D = \frac{m}{2}
  \frac{\hat{\sigma}(A)\hat{\sigma}(B)}{\sigma_{eff}}
\end{equation}

Where all the information on the structure of the hadron in transverse
space is summarized in the value of the scale factor, $\sigma_{eff}$.

The experimental value measured by CDF yields~\cite{Abe97}
\[ \sigma_{eff}=1.45 \pm 1.7^{+1.7}_{-2.3} \mathrm{~mb} \]

It is believed that is largely independent of the center-of-mass
energy of the collision and on the nature of the partonic
interactions.  The experimental evidence is not inconsistent with the
simplest hypothesis of neglecting correlations in momentum fractions.


\section{Cross-Section Results}

Higgs search in the $b\bar{b}h$ channel, however employ cuts on the b-quark
transverse momenta and thus requires theoretical predictions for
exclusive final states. The solution is provided by calculations based
on the parton processes \(q\bar{q},gg \rightarrow b\bar{b}h\).

In this article we evaluate the fully cross section for $b\bar{b}h$
production by requiring that the transverse momentum of both final
state bottom and anti-bottom quarks be larger than 20~GeV. This
corresponds to an experiment measuring the Higgs decay products along
with two high pt bottom quark jets. These cuts reduce the cross
section by one or two orders of magnitude, but also greatly reduce the
background and ensure that the Higgs boson was emitted from a bottom
quark and is therefore proportional to the square of the b-quark
Yakawa coupling.

\begin{figure}[htb]
\includegraphics[width=0.45\textwidth]{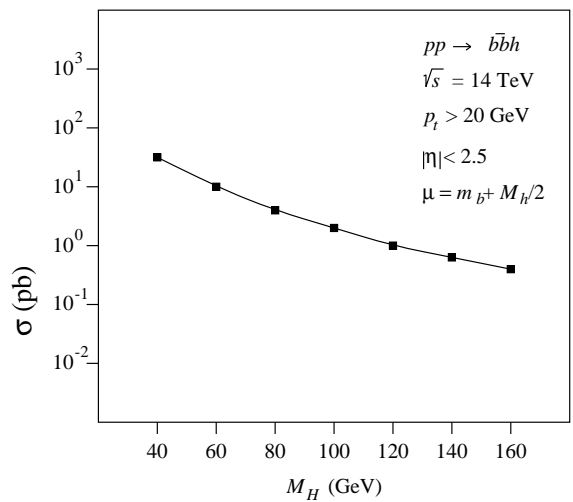}
  \caption{Leading order cross section (pb) for Higgs boson production
    in association with bottom quarks as a function of the Higgs boson
    mass at the LHC.}
  \label{fig:fig_1}
\end{figure}

The cross section for leading order sub-process for Higgs-boson
production in association with bottom quarks obtained using MRST
parton distribution~\cite{Mrt00}, the packages MadGraph~\cite{Stl94}
and HELAS~\cite{Mry92} and the integration was performed by
VEGAS~\cite{Lpg78} as function of Higgs mass for the LHC with
\(\sqrt{s} = 14\)~TeV are displayed in Fig.~\ref{fig:fig_1}.

The cross-section for Higgs production in association with bottom
quarks are not large but may be useful if high luminosity is
available, since the Higgs boson can be ``tagged'' by trigging on the
bottom quarks.

Sizable rates of events where two bottom quarks associate with Higgs
boson are produced contemporarily at the LHC, as a consequence of the
large parton luminosity. The corresponding integrated rate is
evaluated by combining the integrated cross section for Higgs boson
and production at LHC energy.

If one use the cross section for Higgs boson production from \(pp
\rightarrow H+X\), \(\sigma(b\bar{b}) \simeq 5\times10^2\)~$\mu$b as
a value for the scale factor \(\sigma_{eff}=1.45 \)~mb (the observed
value is \(\sigma_{eff}=1.45 \pm 1.7^{+1.7}_{-2.3}\)~mb) one obtain
the cross section for a double parton collision producing a Higgs
boson and a $b\bar{b}$ pair.

The large rate of $b\bar{b}$ pair at the LHC gives rise to a
relatively sizable production of Higgs boson associated with
$b\bar{b}$ quarks.

\begin{figure}[htb]
\includegraphics[width=0.45\textwidth]{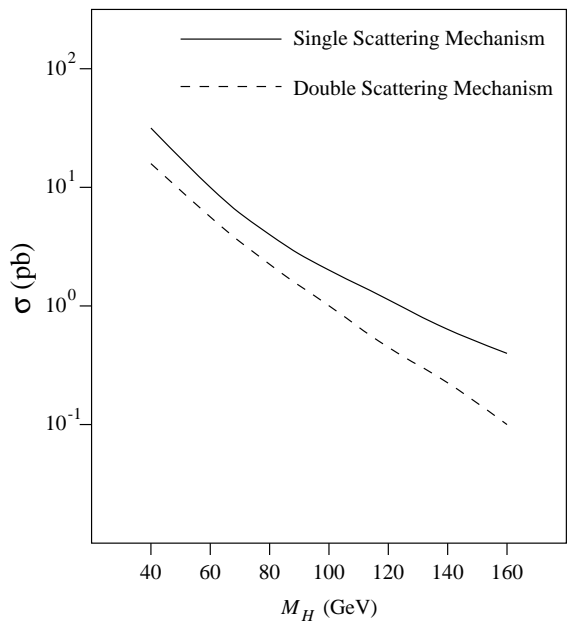}
  \caption{Total cross section for Higgs boson associated with bottom
    quarks in the SM as a function of the Higgs boson mass at LHC
    energy.}
  \label{fig:fig_2}
\end{figure}

Figure~\ref{fig:fig_2} shows the double parton scattering to the Higgs
boson associated with $b\bar{b}$ quarks.

\section{Summary}

In this work we have investigated $b\bar{b}h$ production at the LHC,
which is important discovery channel for Higgs boson in the SM and its
extension in the MSSM at large values of $\tan \beta$, where the
bottom Yakawa coupling is strongly enhanced.

Our calculations corresponds to the cross section for Higgs boson in
association with two tagged b jets in single and double parton
scattering mechanism.

Although the double parton collision cross
section is not large, but it should be taken in consideration because
a sizable rate of events where pairs of quarks are produced at the
LHC, as a consequence of the large parton luminosity.

However, in the hadron collider environment, the large QCD backgrounds
may cause the observation impossible for those events if Higgs bosons
decay hadronically. Individual channels with hadronic decays should be
studied on case by case.





\end{document}